\documentclass[aps,prd,reprint,superscriptaddress,showpacs]{revtex4-1}

\usepackage{graphicx}
\usepackage{dcolumn}
\usepackage{bm}
\usepackage{amsfonts}
\usepackage{amsmath}

\begin{document}

\preprint{APS/123-QED}

\title{Towards Spectral Geometric Methods for Euclidean Quantum Gravity}

\author{Mikhail Panine}
\affiliation{Department of Applied Mathematics, University of Waterloo, Waterloo, Ontario, Canada N2L 3G1}
\author{Achim Kempf}%
\affiliation{Department of Applied Mathematics, University of Waterloo, Waterloo, Ontario, Canada N2L 3G1}
\affiliation{Department of Physics, University of Waterloo, Waterloo, Ontario, Canada N2L 3G1}


\begin{abstract}
The unification of general relativity with quantum theory will also require a coming together of the two quite different mathematical languages of general relativity and quantum theory, i.e., of differential geometry and functional analysis respectively. Of particular interest in this regard is the field of spectral geometry, which studies to which extent the shape of a Riemannian manifold is describable in terms of the spectra of differential operators defined on the manifold. Spectral geometry is hard because it is highly nonlinear, but linearized spectral geometry, i.e., the task to determine small shape changes from small spectral changes, is much more tractable, and may be iterated to approximate the full problem. 
Here, we generalize this approach, allowing, in particular, non-equal finite numbers of shape and spectral degrees of freedom. This allows us to study how well the shape degrees of freedom are encoded in the eigenvalues. We apply this strategy numerically to a class of planar domains and find that the reconstruction of small shape changes from small spectral changes is possible if enough eigenvalues are used. While isospectral non-isometric shapes are known to exist, we find evidence that generically shaped isospectral non-isometric shapes, if existing, are exceedingly rare.

\pacs{04.60.-m, 02.30.Zz, 02.40.-k}
\end{abstract}

\maketitle

\section{Introduction}

A fundamental difficulty with the quantization of general relativity is to separate in the metric the true degrees of freedom from spurious degrees of freedom that merely express choices of coordinates \cite{gibbons1993euclidean,rovelli2004quantum,kiefer2007quantum}. It is of interest, therefore, to obtain a description of curved manifolds in terms of coordinate system independent quantities. Such a set of invariants could be provided by the spectra of canonical differential operators  such as the Dirac operator or Laplacians, at least in the case of Euclidean gravity \cite{kempf2010spacetime,QGQC,landi1997general}. This approach is interesting also because by relating curvature to spectra it naturally translates between the differential geometric language of general relativity and the functional analytic language of quantum theory \cite{kempf2010spacetime}. 

The mathematical discipline concerned with the relationship between the curvature or `shape' of a manifold and the spectra of operators defined on it is known as spectral geometry. Its origins trace back to Weyl \cite{weyl1911asymptotische} and predate quantum mechanics. Concretely,  spectral geometry asks, to what extent properties of manifolds, such as their curvature and their boundaries (or also their boundary conditions) can be determined by spectra of operators on them. The case of the detection of the boundaries of a manifold from spectra has been popularized by Kac's paper entitled ``Can one hear the shape of a drum?" \cite{kac.drum}. Kac's question has inspired investigations into acoustical engineering applications of spectral geometry, see, e.g., the recent \cite{bharaj2015computational}. For simplicity, we will refer to both, the ``hearing" of curvature and the ``hearing" of boundaries as the quest to detect the ``shape" of a manifold from spectra. 

Indeed, it has been shown that, within suitable classes of Riemannian manifolds, it is possible to determine the curvature and / or the shape of the boundary of a (sometimes assumed flat) manifold from spectra. A survey of such results can be found in \cite{datchev2011inverse} and see  also \cite{zelditch1998revolution,zelditch2009inverse,hezari2010inverse}. Examples are reflection-symmetric domains in $\mathbb{R}^{n}$ and surfaces of revolution.

If spectral geometry is to help with the quantization of gravity, i.e., if eigenvalues are to serve as the dynamical degrees of freedom of curvature, then the key question is under which circumstances the spectra of suitable differential operators can fully describe the curvature of a manifold. 

Indeed, there exist classes of manifolds that contain non-isometric but isospectral manifolds, i.e., there are circumstances in which the shape cannot be uniquely determined from spectra. There are techniques for constructing such isospectral non-isometric manifolds, see \cite{gordon2000survey}. These techniques apply only in special instances, however, and it remains unknown how prevalent isospectral nonisometric manifolds are among the generic manifolds that are of interest in physics.  A key question, therefore, has remained open, namely whether the relationship between shape and spectra is in the generic case unique or ambiguous. Are isospectral nonisometric manifolds the norm or the exception? 

The reason why this question has been difficult to answer is that the map from the curvature or shape of a manifold to its spectrum is highly nonlinear, which makes it hard to study its invertibility properties. To make this problem tractable, our approach here is to linearize the problem by applying perturbation theory, so that we can then at least address the question of local invertibility: we ask if knowledge of a small change of spectrum suffices to reconstruct the small change of shape that caused it. Locally inverting the map between shape and spectrum then becomes a question of (pseudo-) inverting a linear operator. If this is possible, i.e., if it is possible to uniquely infer small shape changes from small spectral changes, then the aim is to iterate these infinitesimal steps to obtain finite shape changes from finite differences in the spectrum. That should enable one to then address the original question about the overall uniqueness or ambiguity of shapes for given spectra. 
To summarize, the point of using the linearization of the map from shapes to spectra is that it is local and easier to invert than the full map. We thus trade a global inverse that one cannot construct, and that may not even exist, for a local (pseudo-)inverse that one can construct. 

As a concrete example of this approach, we here study the case of domains in the plane whose shape can be described by a finite number of degrees of freedom. We then use finite element methods to compute the spectra of those domains. Our  observation is  that it is almost always possible to locally determine shape from spectrum, i.e., to determine small shape changes from small spectral changes. Indeed, pairs of isospectral yet non-isometric manifolds appear to be of measure zero.

\section{Infinitesimal Spectral Geometry}

Most difficulties of spectral geometry can be ascribed to the fact that the map between shape and spectrum is highly nonlinear. A tried and true strategy to simplify such problems is to locally linearize them. In the context of inverse spectral geometry, this amounts to trying to determine changes in shape from changes in spectra in a small (infinitesimal) neighborhood of some reference shape. By iterating (similar to integrating) such steps, one may even obtain finite changes in both shape and spectrum. That is, given an initial shape $A$ and a target shape $B$ one can deform $A$ in small (infinitesimal) steps that take its spectrum closer and closer to that of $B$. We dub any such approaches infinitesimal inverse spectral geometry (IISG). Such an approach was used previously in \cite{aasen2013shape} for the spectra of a graph Laplacians on a special family of graphs.

Here, our aim is to develop IISG in a setting that is suitable for numerical investigations. We will use the terminology that the word \it shape \rm  shall denote a Riemannian manifold picked from a suitable class, $\mathcal{G}$. Here, $\mathcal{G}$ can contain shapes that are equivalent by isometry. The set of isometry equivalence classes of $\mathcal{G}$ shall be denoted $[\mathcal{G}]$. 
We will assume that $\mathcal{G}$ can be parametrized in a well-behaved way by $\mathbb{R}^{M}$. We will call the $M$ coordinates in $\mathbb{R}^M$ the shape degrees of freedom. For brevity we will also refer to the points in $\mathbb{R}^{M}$ as shapes. The space of shapes $\mathcal{G}$ will be equipped with a metric $d_{\mathcal{G}}(\cdot,\cdot)$. This then allows us to verify if shapes match up to some predetermined finite threshold $\varepsilon_{\mathcal{G}}$, as required by the inherent limitations of numerical methods. 
Under additional mild conditions one can obtain a metric $d_{[\mathcal{G}]}(\cdot,\cdot)$ on $[\mathcal{G}]$ by taking $d_{[\mathcal{G}]}([A],[B])=\text{inf}_{A \in [A], B\in [B]}d_{\mathcal{G}}(A,B)$.

Given a shape in $\mathcal{G}$ one can compute a finite number $N$ of lowest eigenvalues of its Laplacian. Let $\mathbb{R}^{N}$ be the space of spectra and suppose that it is equipped with a metric $d_{\sigma}(\cdot,\cdot)$. This metric is used to verify if the spectra are equal up to some fixed finite threshold $\varepsilon_{\sigma}$. Moreover, the spectral map must be continuous and differentiable with respect to this metric. The usual Euclidean metric suffices for this task, so it is the one we shall employ from now on. One can construct a spectral map $\sigma:\mathbb{R}^{M} \rightarrow \mathbb{R}^{N}$ between the shape and spectral degrees of freedom.  Since all the studied spectra come from some shape in $\mathcal{G}$, the notation for the spectral distance can be simplified by writing $d_{\sigma}(A,B)$ instead of $d_{\sigma}(\sigma(A),\sigma(B))$. Similarly, one can consider variants of the spectral map that use the spectrum of the Green's operator (which is the spectrum used in what follows) or any other function of the Laplacian. The study of IISG is then reduced to the study of the map $\sigma$.

Given $A,B\in \mathbb{R}^{M}$, IISG strives to construct a parametrized path $P(t)$ in $\mathbb{R}^{M}$ starting at $A$ and ending at $B$. This path must be constructed only with the knowledge of the desired spectrum $\sigma(B)$ and the behaviour of $\sigma$ in the neighborhood of the current point $P(t)$. Since we can not \textit{a priori} guarantee that the target spectrum will be reached every time, we impose the milder condition that $d_{\sigma}(\sigma(P(t)),\sigma(B))$ must be non-increasing along the path. This ensures that even if the method used to construct the path is imperfect and fails to reach a shape with the desired spectrum, the resulting shape is at worst as far from the target as the starting shape was. The simplest way to achieve this is to use a gradient descent optimization method on the spectral distance. This is also the approach used in \cite{aasen2013shape} to study the infinitesimal inverse spectral geometry of Laplacians on graphs.
\begin{equation}\label{GradientMethod}
\frac{d}{dt}P(t) =  - \text{grad} (d_{\sigma}(P(t),B))
\end{equation}

\noindent While simple, this approach presents a number of disadvantages. First, straightforward numerical implementations of the gradient descent are prone to meandering (see \cite{qian1999momentum}, among many others). More importantly though, the conceptual meaning of gradient descent optimization strays from the intuitive idea of locally inverting the spectral map. An improved version of this method can be obtained as follows. Let $v_{\sigma}=\sigma(B)-\sigma(P(t))$ be the desired spectral direction and let $\mathcal{J}(t)$ be the Jacobian matrix of $\sigma$ at $P(t)$. It is easy to show that if the spectral distance $d_{\sigma}(\cdot,\cdot)$ is chosen to be the standard Euclidean metric on $\mathbb{R}^{n}$, the path $P(t)$ defined by the gradient descent method of Equation \eqref{GradientMethod} is equivalent to the path $P(t)$ solving the following equation, up to a reparametrization of $t$:
\begin{equation}\label{GradientMethod2}
\frac{d}{dt}P(t) =  - \frac{1}{2}\text{grad} (\|v_{\sigma}\|^{2}) = \mathcal{J}^{T}(t) v_{\sigma}
\end{equation}

This form suggests the following improvement. Let $\mathcal{J}(t)^{+}$ denote the pseudoinverse of the Jacobian and set:
\begin{equation}\label{PseudoinverseMethod}
\frac{d}{dt}P(t) =  \mathcal{J}^{+}(t) v_{\sigma}
\end{equation}

\noindent In a sense, the pseudoinverse and gradient descent approaches are unified by equations \eqref{GradientMethod2} and \eqref{PseudoinverseMethod}. Most importantly, these equations show that the pseudoinverse method provides an optimal approximation to the local inverse of the spectral map, unlike the gradient method. Indeed, since the pseudoinverse encodes the solution of a least-squares problem, $\sigma(P(t))$ will locally evolve towards $\sigma(B)$ in a way as close as possible to a straight line in the sense of the Euclidean metric in the space of spectra $\mathbb{R}^{N}$. 

Moreover, the pseudoinverse provides a useful canonical generalization of the inverse of a linear map which is applicable to maps also between vector spaces of different dimensions. It allows one to study situations where the numbers of shape and spectral degrees of freedom ($M$ and $N$) are not equal. In such situations the usual inverse function theorem is not applicable. However, the pseudoinverse approach can still give information about how well the spectral degrees of freedom locally encode the shape degrees of freedom. This is a significant technical advance compared to the results of the related study concerning the spectral geometry of graphs reported in \cite{aasen2013shape}, where, by construction, the number of shape degrees of freedom always matched the number of spectral degrees of freedom.

We should note that the partial derivatives of the eigenvalues with respect to the degrees of freedom of shape may be undefined at shapes whose Laplacians have degenerate spectrum. While at such points in shape space, equation \eqref{PseudoinverseMethod} does not hold, fortunately, such cases are not generic. Indeed, it is known that, on a fixed differentiable manifold, metrics that induce Laplacians with nondegenerate spectrum form a residual set in the space of all smooth metrics \cite{bando1983generic}. For the present paper, we restrict our analysis to such generic cases.

\section{Numerical Setup}
Equation \eqref{PseudoinverseMethod} is numerically integrated by the Euler method with a variable step size. If the step results in a shape whose spectrum is closer to the target one, the step size is increased. Otherwise, the step is cancelled and the step size is decreased. The increase and decrease of step sizes are accomplished by multiplying the step size by constant factors $1.1$ and $0.7$, respectively. The precise choice of those values is of course arbitrary. Increasing the step size speeds up the execution of the algorithm, which can be quite lengthy if the distance $d_{[\mathcal{G}]}(A,B)$ is large. The decrease in step size helps the algorithm converge. The algorithm is stopped when either $\sigma(P(t))$ becomes close enough to $\sigma(B)$, as dictated by the tolerance $\varepsilon_{\sigma}$, or the step size becomes negligibly small, indicating that the algorithm is stuck. The distance $d_{[\mathcal{G}]}(P(t),B)$ is then computed and if it is smaller than the tolerance threshold $\varepsilon_{\mathcal{G}}$ it is deemed that the algorithm has succeeded. Otherwise, the run is deemed a failure. By generating random pairs $(A,B)$ one can test how the success rate depends on various factors such as the number of considered eigenvalues and the initial distance $d_{[\mathcal{G}]}(A,B)$ between shapes. Of particular interest will be the question whether the rate of success approaches $1$ as $d_{[\mathcal{G}]}(A,B)$ is decreased, provided that $N$ is large enough, as compared to the number of shape degrees of freedom $M$.

We apply this method to a class of domains in $\mathbb{R}^{2}$ equipped with the standard Laplacian with Dirichlet boundary conditions. The spectra were computed using the FreeFem++ \cite{freefem} finite element solver. The boundary of the studied domains is given by a radial function $r(\phi)$ of the standard polar angle $\phi$:

\begin{equation} \label{Radius}
r(\phi)=a+b \exp \left(C_{0} + \sum_{k=1}^{\frac{M-1}{2}} \left[ C_{k} \cos(k\phi)+S_{k} \sin(k \phi) \right] \right)
\end{equation}

\begin{figure}
\includegraphics[width=7cm]{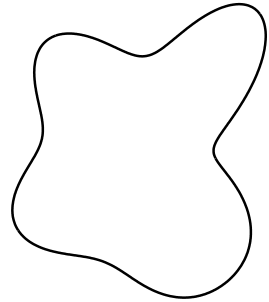}
\caption[Example Shape.]{An example shape for $M=11$.}
\label{ExampleShape}
\end{figure}

\noindent The arbitrary positive parameters $a$ and $b$ were set to $a=0.1$, $b=0.9$. The purpose of $a$ is to set a small but finite minimal radius to the studied shapes, a technical condition that ensures proper functioning of the finite element solver. The value of $b$ is set so that $r(\phi)=1$ if all of the Fourier coefficients $C_{i}$ and $S_{i}$ vanish. The space of shape degrees of freedom is taken to be the space of the first $M$ Fourier coefficients. The studied range of the coefficients yields shapes with diameter roughly between $1$ and $10$. The shape tolerance threshold is set to be $\varepsilon_{\mathcal{G}} = 0.005$, which is well smaller than the typical size of the studied shapes. The spectral threshold is set to $\varepsilon_{\sigma} = \sqrt{10^{-9}} \approx 3.16 \cdot 10^{-5}$, a number that was chosen to be compatible with the threshold $\varepsilon_{\mathcal{G}}$, as will be discussed in more detail below. A example shape is shown in Figure \ref{ExampleShape}.

A metric $d_{\mathcal{G}}(\cdot,\cdot)$ on this class of shapes is obtained by using the Hausdorff distance between the boundaries viewed as subsets of $\mathbb{R}^{2}$ \cite{munkres2000topology}. A corresponding notion of distance $d_{[\mathcal{G}]}(\cdot,\cdot)$  on $[\mathcal{G}]$ is obtained by minimizing $d_{\mathcal{G}}(\cdot,\cdot)$ over the isometries of the plane. In practice, this is done by translating both shapes so that their centers of mass coincide with the origin and then minimize the distance over all rotations and reflections of one of the shapes.

\section{Numerical Results}

Rates of success of our approach were obtained by fixing $\frac{M-1}{2}=1...5$ (i.e. $M=3,5,...,11$), generating random pairs $(A,B)$ and running the algorithm for $N=1...40$. The number of random pairs is between $1250$ and $1750$, depending on $M$. Random pairs whose initial shape distance was too small, i.e., less than the shape tolerance threshold $\varepsilon_{\mathcal{G}}$ (\textit{i.e.} automatic successes) were excluded. For a fixed $M$, the success rate of the algorithm was analyzed as a function of the isometry-invariant shape distance $d_{[\mathcal{G}]}(A,B)$ and of the number $N$ of considered eigenvalues.

\begin{figure}
\includegraphics[width=7cm]{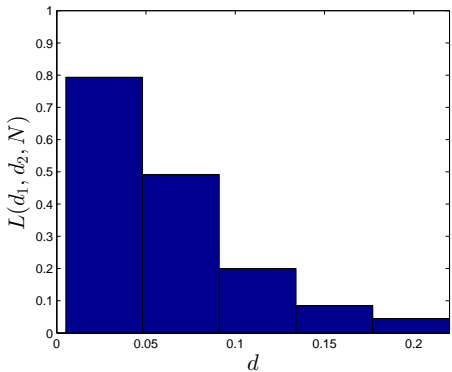}
\caption[Success Rate Estimate.]{Success rate as a function of initial shape distance for $M=11$ and $N=40$.}
\label{BinnedSuccess}
\end{figure}

In our analysis we employ two ways to represent the dependence of the success rate on the initial shape distance. The first is to present $L(d_{1},d_{2},N)$ which is the proportion of all pairs $(A,B)$ with $d_{1} < d_{[\mathcal{G}]}(A,B) \leq d_{2}$ for which the algorithm succeeds for a given $N$. Thus, it is simply the intuitive notion of success rate in a bin $(d_{1},d_{2}]$. This success rate is illustrated on Figure \ref{BinnedSuccess} for $M=11$ and $N=40$. Notice that the success rate decays rapidly with $d_{[\mathcal{G}]}(A,B)$. Still, at short distances, the success rate is quite encouraging.

\begin{figure}
\includegraphics[width=7cm]{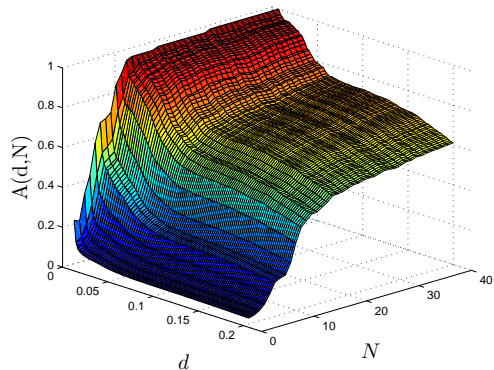}
\caption[Success Rate Estimate.]{Success rate for runs with initial shape distance less than $d$ for $M=11$ and varying $N$.}
\label{LimitSuccess}
\end{figure}

In order to probe the short distance behaviour more closely, we use a second, specialized way to represent the dependence of the  success rate on the initial shape distance. Let $R(d)$ denote the number of pairs $(A,B)$ such that $d_{[\mathcal{G}]}(A,B) \leq d$ and let $S(d,N)$ denote the number of pairs $(A,B)$ such that the algorithm has succeeded in finding a shape isometric to $B$ when using $N$ eigenvalues. We then define the accumulated success rate $A(d,N)=S(d,N)/R(d)$. This allows us to better represent the success rate as $d_{[\mathcal{G}]}(A,B)$ goes to zero, as $A(d,N)$ is the success rate in a bin $[0,d]$. For $M=11$ and $N=1...40$, $A(d,N)$ is illustrated on Figure \ref{LimitSuccess}. Two key features become apparent. First, the success rate of the algorithm indeed 
tends towards $1$ as the distance between the starting and target shape goes to zero, provided that $N$ is large enough. This indicates a success of the program of infinitesimal inverse spectral geometry. The second feature is the threshold at which $N$ becomes large enough. Notice that the success rate is rapidly increasing as $N$ goes from $1$ to roughly $M$, and then plateaus near $1$. This indicates that, at least for the considered space of shapes, it is sufficient to attain an approximate match between $N$ and $M$ to be able to reconstruct infinitesimal changes in shape from infinitesimal changes in spectrum. The reason this match is approximate is that the description of the space of shapes possesses redundancy. Indeed, since rotations about the origin are isometries, one shape degree of freedom is pure gauge. This suggests the possibility that the limit $M, N \rightarrow \infty$ of infinitely many shape degrees of freedom and full spectrum could be successfully treated by the same approach as the finite dimensional case, although we will not pursue this limit here.

Let us now investigate the cases where the algorithm does not succeed. We start by considering the possible outcomes of the algorithm. The four possibilities are: 1) both shapes and spectra match, 2) shapes match but not spectra, 3) spectra match but not shapes and 4) neither match. The first case is of course that of the algorithm succeeding and warrants no further explanation.

The second case is an unavoidable artifact of finite numerical precision and of the fact that
identical balls in shape space will correspond to domains in spectral space of significantly varying volume and shape. It is thus not possible to relate the size of those domains by the choice of two constant thresholds. In other words, having picked a constant $\varepsilon_{\mathcal{G}}$, one can not pick a constant $\varepsilon_{\sigma}$ such that isometry up to the threshold $\varepsilon_{\mathcal{G}}$ always implies isospectrality up to the the threshold $\varepsilon_{\sigma}$. We sidestep this issue by, conservatively, counting those ambiguous cases as failures. Thus, the success rates that we report are lower bounds for the success rates for the chosen spectral and shape tolerances.

\begin{figure}
\includegraphics[width=7cm]{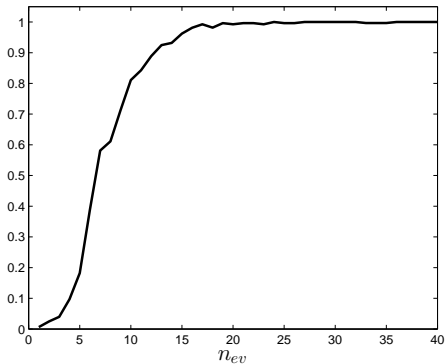}
\caption[Proportion of Isometric runs in Isospectral ones.]{Proportion of isometric runs among isospectral ones for $M=11$.}
\label{ProportionIsometricIsospectral}
\end{figure}

The third possible outcome is interesting, as it corresponds to domains that are non-isometric but are isospectral on their first $N$ eigenvalues. That is, they are potential counterexamples to the general program of infinitesimal inverse spectral geometry. Figure \ref{ProportionIsometricIsospectral} illustrates the proportion of isometric runs among isospectral ones for varying $N$. Notice that this proportion goes to $1$ as $N$ increases. This indicates that, at least for the studied space of shapes, non-isometric isospectral shapes form a set of measure zero.

Finally, the fourth outcome is the algorithm getting stuck when the  right hand side of Equation \eqref{PseudoinverseMethod} vanishes. Since the transpose and the pseudoinverse of a matrix share their kernels, $\mathcal{J}^{+}(t) v_{\sigma}=0$ if and only if $\mathcal{J}^{T}(t) v_{\sigma}=0$. It is straightforward to show that $\mathcal{J}^{T}(t) v_{\sigma}=0$ if and only if $P(t)$ is a critical point of $d_{\sigma}(\cdot,\sigma(B))$. That is, our minimization algorithm gets stuck in a local minimum, as minimization algorithms are prone to do. It could be useful, therefore, to employ more sophisticated methods for overcoming trapping in local minima.

\section{Conclusions and Outlook}
It is of great interest for the quantization of gravity to be able to separate the true degrees of freedom of a (pseudo-) Riemannian manifold (i.e., the curvature of both the bulk and/or the boundary) from the spurious degrees of freedom in the metric that merely express choices of coordinates.   

To this end, we started with the observation that the eigenvalues of Laplacians and other natural differential operators on the manifold are true degrees of freedom of the metric because they are geometric invariants, i.e., they are independent of diffeomorphisms. The key question, however, is whether these eigenvalues encode \it all \rm of the true degrees of freedom. Answering this question is complicated by the fact that the relationship between the metric and the associated eigenvalues is nonlinear. For this reason, we introduced a perturbative approach, namely we re-examined the task of reconstructing the curvature of a boundary from knowledge of the eigenvalues by turning the usual question {\it ``Can one hear the shape of a drum?"} into the much more manageable local question {\it ``Does a small change of sound tell the corresponding small change of shape?"}. This method of ``infinitesimal" inverse spectral geometry allowed us to study the invertibility properties of the nonlinear map from shapes to spectra both a) in the neighborhood of a shape and, b) to some extent also in the nonlinear regime via iteration: 

(a) We considered the ``limit" behavior for decreasing distances between the initial and target shape for varying numbers $M,N$ of shape and spectral degrees of freedom respectively. We found that as soon as the number $N$ of spectral degrees of freedom matches or exceeds the threshold of $M$ shape degrees of freedom the success rate tends towards $100\%$. 
This confirms analytic expectations on the basis of linearizing the nonlinear map from $\mathbb{R}^{M}$ to $\mathbb{R}^{N}$. Note that this reasoning applies equally to the case of a curved manifold whose metric is parametrized by $\mathbb{R}^{M}$. The fact that the threshold is indeed $M$, rather than some function of $M$ is encouraging regarding the limit $M,N \rightarrow \infty$. 

(b) We investigated cases of considerable shape distance by iterating the small steps. In particular, we focused on those runs of
the algorithm that found a final shape with the desired target spectrum. We found that as the number, $N$, of considered eigenvalues is increased beyond the number of shape degrees of freedom $M$, those runs that also found the desired shape became dominant. In fact, the proportion of such cases seems to rapidly go to $1$ as $N$ increases. This strongly suggests that counterexamples to inverse spectral geometry are of measure zero or even absent within the considered class of shapes.
This is consistent with the fact that, to the best of our knowledge, all counterexamples to the spectral geometry program in the plane \cite{gordon1996you,cannothear,gordon1992isospectral} are not in our class of shapes because they are non-star-shaped domains with non-smooth boundaries. The counterexamples are thus not only not part of the set of shapes we studied, but cannot even be approximated by shapes from the considered set, no matter how many Fourier coefficients are used. It will therefore be interesting to further study the properties of this class of shapes using, in particular, methods inspired by those used in, for example,  \cite{zelditch1998revolution,zelditch2009inverse,hezari2010inverse}. There, the proofs rely on the fact that the boundaries are analytic functions, which is also true in our case.

Returning to our original motivation, in order to apply our infinitesimal inverse spectral geometry technique to the quantization of  (Euclidean) gravity, it will be necessary to generalize our new pseudoinverse-based method, see equation \eqref{PseudoinverseMethod}, to the degrees of freedom of the metric on the bulk of a compact manifold. Ultimately, this will involve developing a functional analytic generalization of the pseudoinverse-based method in order to handle infinitely many shape and spectral degrees of freedom.

\begin{acknowledgments}
 AK and MP acknowledge support through the Discovery, NSERC-CGS-M and NSERC-PGS-D (Canada Graduate Scholarship) programs of the Natural Sciences and Engineering Research Council (NSERC) of Canada. MP also acknowledges support through the Ontario Graduate Scholarhip (OGS) program. This work was made possible by the facilities of the Shared Hierarchical Academic Research Computing Network (SHARCNET) and Compute/Calcul Canada.
 \end{acknowledgments}


\phantomsection  
\addcontentsline{toc}{section}{\textbf{References}}

\bibliography{GenericBibliography}

\end{document}